\documentclass[12pt,twoside,onecolumn,draftcls]{IEEEtran} %,draft Specifies the documentclass
\usepackage{graphicx,cite,amsmath,amssymb,hhline,booktabs,stfloats}
\usepackage{verbatim}
\usepackage[utf8]{inputenc}
\usepackage{subfigure}

\begin{document}

\title{Novel energy detection using uniform noise distribution}

%\iffalse

\author{
\bigskip
\medskip {\normalsize $\mbox{Kezhi Wang}^{}$, {\em Student Member, IEEE,} $\mbox{Yunfei Chen}^{}$, {\em Senior Member, IEEE,}
    and $\mbox{Jiming Chen}^{}$, {\em Senior Member, IEEE}}
\thanks{K. Wang and Y. Chen are with the School of Engineering, University of Warwick, CV4 7AL
Coventry, U.K. (e-mails: \{Kezhi.Wang, Yunfei.Chen\}@warwick.ac.uk).}
\thanks{J. Chen is with Dept. of Control Science and Engineering,
Zhejiang University, Hangzhou 310027, China.}
\thanks{This paper has been submitted to Wireless Communications and Mobile Computing on 18-Oct-2013, and it is under second round of reviewing.}
}

\maketitle
% Produces the title.

\renewcommand{\baselinestretch}{1.73} \normalsize

\begin{abstract}
Energy detection is widely used in cognitive radio due to its low
complexity. One fundamental challenge is that its performance
degrades in the presence of noise uncertainty, which inevitably
occurs in practical implementations. In this work, three novel
detectors based on uniformly distributed noise uncertainty as the
worst-case scenario are proposed. Numerical results show that the
new detectors outperform the conventional energy detector with
considerable performance gains.
\end{abstract}

\begin{keywords}
Average likelihood ratio test; cognitive radio; energy
detection; noise uncertainty
\end{keywords}

\section{Introduction}
Cognitive radio is considered to be a solution to the problems of
spectrum under-utilization and ``spectrum scarcity'' \cite{Haykin}.
% You must have at least 2 lines in the paragraph with the drop letter
% (should never be an issue)Spectrum sensing algorithms in Cognitive Radio can
The IEEE 802.22 Working Group has developed a standard for wireless
regional area networks (WRAN) based on cognitive radio, which can
operate on unused digital TV broadcast bands. In order to avoid
interfering the primary services, the main task of WRAN is to detect
the presence of the primary users. Many spectrum sensing methods
have been proposed in the literature which can be mainly divided
into three types: energy detection \cite{Urkowitz}, matched-filter
detection \cite{Kay} and feature detection \cite{Cabric}. \iffalse
Also, feature detection mainly includes cyclostationary
\cite{81007}, eigenvalue-based \cite{5089517,5397901} and
covariance-based detection \cite{5425309,5604095}. Also, feature
detection mainly includes cyclostationary, eigenvalue-based and
covariance-based detection. Among them, matched-filter needs to know
the waveforms and channels of the primary users, cyclostationary
detection requires knowledge of cyclic frequencies of the primary
users, eigenvalue-based detection and covariance-based detection
need covariance matrix of the signals received at the secondary
users, which may not be available or computationally complicated. On
the other hand, A performance comparison between feature detector
and energy detector can be found in \cite{5089517}. In general,
feature detection outperforms energy detection at the cost of much
higher complexity. This work focuses on energy detection. \fi Among
them, energy detection does not need any information about the
primary signals and is widely used due to its simplicity
\cite{Urkowitz}. Most works on energy detection assume that the
noise power is accurately known. In reality, it is very difficult to
obtain the accurate value of the noise power, leading to noise
uncertainty \cite{Sonnenschein}. The noise uncertainty can severely
degrade the performance of energy detection \cite{Tandra}.

In this work, three new energy detection schemes are proposed by
using the distribution of the noise power in energy detection to
remove the need for the noise power in the detector such that noise
uncertainty can be avoided and energy detection can be improved. To
do this, uniformly distributed noise power is adopted as the
worst-case scenario, as the value of the noise power is equally
likely across a certain interval. This uniform
distribution is for noise power, not for noise uncertainty.
%Uniform distribution of noise power is also assumed in some other papers %\cite{uni2,uni1,uni3,uni4,uni5,uni6,uni7}.

Numerical results show that the proposed new schemes have better
performances than the conventional energy detector with noise
uncertainty. They also show that even when there is a mismatch
between the assumed uniform distribution and the actual distribution
of the noise power, the new schemes still have considerable
performance gains, verifying the robustness of the new schemes.
Although the average likelihood ratio test (ALRT) \cite{904013} is
not a new method and has been applied in many other works, the
detectors from it are new and represent novelty. Due to the limited
space, only the most relevant references on ALRT are discussed here,
although there are other less relevant references on feature
detectors.

\section{System Model}
Consider the binary hypothesis testing problem for energy detection
as
\begin{equation}\label{eq1}
\begin{matrix}
H_{0}: &x[i]=w[i] \\
H_{1}: &x[i]=s[i]+w[i]
\end{matrix}
\end{equation}
where $H_{0}$ represents the hypothesis that the signal is absent,
$H_{1}$ indicates the hypothesis that the signal is present,
$i=1,2,...,N$ index the $N$ signal samples, $s[i]$ is the Gaussian
signal with mean zero and variance $\beta^{2}$ \cite[pp. 142]{Kay},
and $w[i]$ is the additive white Gaussian noise with mean zero
and variance $\sigma^{2}$. This assumption of Gaussian signal is widely used in previous works \cite{Kay}. %6484995
The Neyman-Pearson (NP) rule is commonly used. The performance of NP
detection is measured by the pair of the detection probability $P_d$
and the false alarm probability $P_{fa}$. Similar to
\cite{Urkowitz}-\cite{Cabric}, this work does not consider the
traffic load, which is the case when the primary user has very light
traffic.

From (\ref{eq1}), one can get the probability density function (PDF)
of $x[i]$ under $H_1$ as
\begin{equation}\label{eq3}
f(x[i]| H_1,\sigma ^2)=\frac{1}{\sqrt{2 \pi \left(\beta ^2+\sigma
^2\right)}}e^{-\frac{x[i]^2}{2 \left(\beta ^2+\sigma ^2\right)}}
\end{equation}
and the PDF of $x[i]$ under $H_0$ as
\begin{equation}\label{eq4}
f(x[i]|H_0,\sigma ^2)=\frac{1}{\sqrt{2 \pi  \sigma
^2}}e^{-\frac{x[i]^2}{2 \sigma ^2}}.
\end{equation}
Then, $X\sim \mathcal{N}( \textbf {0}, \sigma\textbf {I}$) under
$H_0$ and $X\sim \mathcal{N}( \textbf {0}, (\sigma^2+\beta^2)\textbf
{I}$) under $H_1$. The likelihood ratio test can be constructed
according to \cite[eq. (5.1)]{Kay} as
\begin{equation}\label{eqq4}
L_1'(X)=\frac{f\left(X|H_1,\sigma ^2\right)}{f\left(X|H_0,\sigma
^2\right)}\;\; \overset{H_1}{\underset{H_0}{_{<}^{>}}} \;\;
\gamma_1'
\end{equation}
where $\gamma_1'$ is the detection threshold and
\begin{equation}\label{eq6}
\begin{aligned}
f(X|H_1,\sigma ^2)&=\frac{1}{\left[2 \pi \left(\beta ^2+\sigma
^2\right)\right]^{N/2}} e^{-\frac{\sum _{x=1}^N x[i]^2}{2
\left(\beta ^2+\sigma ^2\right)}},
\end{aligned}
\end{equation}
\begin{equation}\label{eq7}
\begin{aligned}
f(X|H_0,\sigma ^2)&=\frac{1}{\left(2 \pi  \sigma ^2\right)^{N/2}}
e^{-\frac{\sum _{x=1}^N x[i]^2}{2 \sigma ^2}}.
\end{aligned}
\end{equation}
Thus, one has from (\ref{eqq4})
\begin{equation}\label{eqe2}
\begin{aligned}
&L_1(X)=\sum _{i=1}^N x[i]^2\;\;
\overset{H_1}{\underset{H_0}{_{<}^{>}}} \;\; \gamma_1.
\end{aligned}
\end{equation}
Denote the false alarm probability as
\begin{equation}\label{eqe5}
\begin{aligned}
P_{fa}=Pr\{L_1(X)>\gamma_1|H_0\},
\end{aligned}
\end{equation}
and the detection probability as
\begin{equation}\label{eqe6}
\begin{aligned}
P_d=Pr\{L_1(X)>\gamma_1|H_1\}.
\end{aligned}
\end{equation}
Therefore, the threshold $\gamma_1$ can be determined as \cite[pp.
143]{Kay}
\begin{equation}\label{eqq7}
\begin{aligned}
\gamma_1=Q_{{\chi}^2_N }^{-1}(P_{fa},\sigma ^2),
\end{aligned}
\end{equation}
where $Q_{{\chi}^2_N }$ is the right-tail probability for a ${\chi}^2$
random variable with $N$ degrees of freedom and $Q_{{\chi}^2_N
}^{-1}$ is the inverse function of $Q_{{\chi}^2_N }$. This detector
requires knowledge of the noise power $\sigma ^2$ in order to
calculate the detection threshold. In practice, $\sigma ^2$ has to
be estimated and the estimation error is random \cite{Tandra}. As a
result, the estimate of $\sigma ^2$ used in the detection is also a
random variable. This noise uncertainty leads to detection errors in
(\ref{eqe2}). The proposed new detectors will not suffer from this estimation error and thus outperform (\ref{eqe2}).

Using the maximum likelihood method to estimate $\sigma ^2$ with $K$ samples,
the PDF of the estimate is \cite{yunfei4}
\begin{equation}\label{esti}
f(\hat{\sigma ^2})=K\;\frac{2^{-\frac{K}{2}} \sigma^{-K}
(K\;\hat{\sigma ^2})^{\frac{K}{2}-1} e^{-\frac{K\hat{\sigma ^2}}{2
\sigma ^2}}}{\Gamma \left(\frac{K}{2}\right)}.
\end{equation}
where $\hat{\sigma ^2}$ is the estimate of the noise power $\sigma
^2$. Note that this estimator $\hat{\sigma ^2}$ uses pilot symbols
in the training period of the secondary user. Such pilot symbols are
not available from the primary user for spectrum sensing and thus,
matched-filter detection cannot be used. Denote the detector in
(\ref{eqe2}) as the NP-LRT detector, which is the conventional
energy detector using the maximum likelihood estimate of the noise power.

One way of avoiding the noise uncertainty in (\ref{eqe2}) is to
remove the use of $\sigma ^2$ in the detection. This can be achieved
by averaging the likelihood function or the likelihood ratio over
the distribution of $\sigma ^2$ based on the ALRT principle.
References \cite{uni2} and \cite{uni3} analyzed the detector
performance by averaging $P_d$ and $P_{fa}$. They did not average
the decision variable to eliminate the noise uncertainty. We assume
that $\sigma ^2$ is uniformly distributed over a certain interval as
$\sigma^2\epsilon \left (\Delta_{min}, \Delta_{max}  \right )$, with
the PDF of
\begin{equation}\label{pdf}
f\left(\sigma ^2\right)=\frac{1}{\Delta _{max}-\Delta _{min}}.
\end{equation}
Uniform distribution has been widely used as a universal
non-informative prior in many applications \cite{Shulman},
especially when the parameter space is finite but the value and the
distribution are unknown \cite{Jeffreys}. Compared with
other distributions, such as log-normal distribution, % \cite{Jouini},
the PDF of the uniform distribution has a simple structure and therefore closed-form energy detectors can be derived. Also, uniform distribution can be regarded as the worst-case
scenario because the noise power is equally likely anywhere in the
whole interval \cite{BaiErWei}. Thus, this is a very useful
benchmark. In reality, the noise power equals $N_0B$, where $N_0$ is
the single-sided power spectral density and $B$ is the bandwidth.
Further, $N_0=kT$ where $k$ is the Boltzman constant and $T$ is the
temperature. Thus, as long as $B$ is fixed and $T$ is uniformly
distributed over a certain interval with limited low temperature and
high temperature, the noise power is also uniformly distributed in
this case. \iffalse Similar discussions can be found in \cite{uni7}
and the references therein and references citing \cite{uni7}.\fi
Most receivers do have an operating range of temperature, which can
be used to determine $\Delta_{min}$ and $\Delta_{max}$ together with
$B$ and $k$. Note that in realistic situation, one also needs to
consider electrical and thermal noise, frequency response and other
factors, but to simplify the detector, this work only considers the
ideal situation where $N_0=kT$. In the realistic situation, one can
assume the noise power equals $K N_0B$, where $K$ is a constant
which takes electrical noise, frequency influence and other factors
into account. \iffalse To simplify the analytical process, both
external and internal noise power are generalized by using
$\sigma^2$ in this work, although the distribution of external noise
power may not be uniform.\fi

%%%%%%%%%%%%%%%%%%%%%%%%%%%%%%%%%%%%%%%%%%%%%%%%%%%%%%%%%%%%
\section{New energy detectors}
In this section, three new detectors based on the uniform
distribution of $\sigma ^2$ are proposed. The first one is denoted
as NP-AVE detector which averages the likelihood function of each
sample over the uniform distribution of $\sigma^2$. The second one
is denoted as NP-AVN detector which averages the overall likelihood
function of all the samples over the uniform distribution of
$\sigma^2$. It is very difficult to obtain the exact average likelihood ratio
over the uniform distribution of $\sigma^2$. Thus, this work
conducts averaging over the numerator and the denominator separately
to obtain tractable approximate detectors. This is somewhat brute-forced but still useful. The third one is denoted
as the NP-LLR detector, which is obtained by averaging the
log-likelihood ratio over the distribution of $\sigma^2$. Note that
there are no closed-form expressions of $P_d$ and $P_{fa}$ for the
NP-AVE and NP-AVN detectors and one has to calculate them by
numerical integrations. For the NP-LLR detector, the closed-form
expression is available and will be provided.
\subsection{NP-AVE detector}
One can get NP-AVE detector as
\begin{equation}\label{eq13}
\begin{aligned}
&L_2(X)=\prod_{i=1}^N \frac{f(x[i]|H_1)}{ f(x[i]|H_0)}\;\;
\overset{H_1}{\underset{H_0}{_{<}^{>}}}
 \;\; \gamma_2
\end{aligned}
\end{equation}
where $\gamma_2$ is the detection threshold and
\begin{equation}\label{eq10}
\begin{aligned}
&f(x[i]|H_1)=(\text{$\Delta _{max}$}-\text{$\Delta _{min}$})
\left(\sqrt{\frac{2}{\pi } \left(\beta ^2+\text{$\Delta
_{max}$}\right)} e^{-\frac{x[i]^2}{2 \left(\beta ^2+\text{$\Delta
_{max}$}\right)}} -\sqrt{\frac{2}{\pi } \left(\beta ^2+\text{$\Delta
_{min}$}\right)}\text{  }e^{-\frac{x[i]^2}{2 \left(\beta
^2+\text{$\Delta _{min}$}\right)}}\right. \\
&\left.+x[i]\; \text{Erf}\left(\frac{x[i]}{\sqrt{2 \left(\beta
^2+\text{$\Delta _{max}$}\right)}}\right)-x[i]\; \text{Erf}\left(\frac{x[i]}{\sqrt{2 \left(\beta
^2+\text{$\Delta _{min}$}\right)}}\right)\right),
\end{aligned}
\end{equation}
and
\begin{equation}\label{eq12}
\begin{aligned}
&f(x[i]|H_0)=(\text{$\Delta _{max}$}-\text{$\Delta _{min}$})
\left(\sqrt{\frac{2}{\pi } \Delta _{max}}\; e^{-\frac{x[i]^2}{2
\Delta _{max}}} -\sqrt{\frac{2}{\pi }\Delta _{min}}\; e^{-\frac{x[i]^2}{2
\Delta
_{min}}}+x[i]\;\right. \\
&\left. \text{Erf}\left(\frac{x[i]}{\sqrt{2 \Delta _{max}}}\right)-x[i]\; \text{Erf}\left(\frac{x[i]}{\sqrt{2 \Delta
_{min}}}\right)\right).
\end{aligned}
\end{equation}
\noindent \textit{Proof}: See Appendix. $A$.

\noindent Due to the complexity of the decision variable in
(\ref{eq13}), the detection threshold $\gamma_2$ will be calculated
by simulation.

\subsection{NP-AVN detector}
The NP-AVN detector is derived as
\begin{equation}\label{eq19}
\begin{aligned}
&L_3(X)= \left(\left(\beta ^2+\Delta _{max}\right)^{1-\frac{N}{2}}
  EI\left(2-\frac{N}{2},\frac{\sum _{i=1}^N x[i]^2}{2
\left(\beta ^2+\Delta _{max}\right)}\right)-\left(\beta ^2+\Delta
_{min}\right)^{1-N/2}
\right. \\
&\left.\times EI\left(2-\frac{N}{2},\frac{\sum _{i=1}^N x[i]^2 }{2
\left(\beta ^2 +\Delta
_{min}\right)}\right)\right)/\left(\left(\Delta
_{max}\right)^{1-\frac{N}{2}}
 EI\left(2-\frac{N}{2},\frac{\sum
_{i=1}^N x[i]^2}{2 \Delta _{max}}\right) \right. \\
&\left.-\left(\Delta _{min}\right)^{1-\frac{N}{2}} EI
\left(2-\frac{N}{2},\frac{\sum _{i=1}^N x[i]^2}{2 \Delta _{min}
^2}\right)\right)\;\; \overset{H_1}{\underset{H_0}{_{<}^{>}}} \;\;
\gamma_3
\end{aligned}
\end{equation}
where the exponential integral function
$\text{\textit{$EI$}}(\text{\textit{$n$}},\text{\textit{$z$}})=\int
_1^{\infty
}\text{\textit{$e$}}^{-\text{\textit{$z$}}\text{\textit{$t$}}}
/\text{\textit{$t$}}^{\text{\textit{$n$}}}d\text{\textit{$t$}}$
\cite{Smith} and $\gamma_3$ is the detection threshold of the NP-AVN
detector.

\noindent \textit{Proof}: See Appendix. $B$.

\noindent Again, due to the complicated structure of the decision
variable in (\ref{eq19}), $\gamma_3$ has to be calculated via
simulation.
\subsection{NP-LLR detector}
From (\ref{eqe2}), one has
\begin{equation}\label{t4}
\begin{aligned}
&L_4(X)=\sum _{i=1}^N x[i]^2\;\;
\overset{H_1}{\underset{H_0}{_{<}^{>}}} \;\; \gamma_4.
\end{aligned}
\end{equation}
Then, one can get $P_{fa}$ as
\begin{equation}\label{eqe7}
\begin{aligned}
&P_{fa}=\left(2^{1-\frac{N}{2}} \gamma_4^{N/2} \left(\Delta _{min}^{1-\frac{N}{2}} e^{-\frac{\gamma_4}{2 \Delta _{min}}}-\Delta _{max}^{1-\frac{N}{2}} e^{-\frac{\gamma_4}{2 \Delta _{max}}}\right)+(\Delta _{min} (N-2)-\gamma_4) \right. \\
&\left. \Gamma \left(\frac{N}{2},\frac{\gamma_4}{2 \Delta
_{min}}\right)-(\Delta _{max} (N-2)-\gamma_4) \Gamma \left(\frac{N}{2},\frac{\gamma_4}{2 \Delta
_{max}}\right)\right)/\left((N-2) (\Delta _{min}-\Delta _{max})
\right. \\
&\left. \Gamma \left(\frac{N}{2}\right)\right),
\end{aligned}
\end{equation}
and $P_d$ as
\begin{equation}\label{eqe8}
\begin{aligned}
&P_d=\left(2^{1-\frac{N}{2}} \gamma_4^{N/2} \left((\Delta _{min}+\beta^2)^{1-\frac{N}{2}} e^{-\frac{\gamma_4}{2 (\Delta _{min}+\beta^2)}} -(\Delta _{max}+\beta^2)^{1-\frac{N}{2}} e^{-\frac{\gamma_4}{2 (\Delta _{max}+\beta^2)}}\right)\right. \\
&\left.+((N-2) (\Delta _{min}+\beta^2)-\gamma_4) \Gamma \left(\frac{N}{2},\frac{\gamma_4}{2 (\Delta _{min}+\beta^2)}\right)-((N-2) (\Delta _{max}+\beta^2)-\gamma_4) \right. \\
&\left.\times \Gamma \left(\frac{N}{2},\frac{\gamma_4}{2 (\Delta
_{max}+\beta^2)}\right)\right)/\left((N-2) (\Delta _{min}-\Delta
_{max}) \Gamma \left(\frac{N}{2}\right)\right).
\end{aligned}
\end{equation}

\noindent \textit{Proof}: See Appendix. $C$.

\noindent Using (\ref{eqe7}), the detection threshold can be
determined as
\begin{equation}\label{eqe9}
\begin{aligned}
\gamma_4=P_{fa}^{-1}(\Delta _{min},\Delta _{max}).
\end{aligned}
\end{equation}
where $P_{fa}^{-1}(\cdot, \cdot)$ is the inverse function of
$P_{fa}(\cdot, \cdot)$ with parameters $\Delta _{min}$ and $\Delta
_{max}$. Denote the detector in (\ref{t4}) as Neyman-Pearson log-likelihood ratio
(NP-LLR) detector. The receiver operating characteristic (ROC) curve
for NP-LLR detector can be easily obtained using (\ref{eqe9}) in
(\ref{eqe8}). Note that (\ref{t4}) has the same decision variable as
the conventional detector in (\ref{eqe2}). However, the detection
threshold in (\ref{eqe2}) depends on the noise power and thus
suffers from noise uncertainty, while the detection threshold in
(\ref{eqe9}) is independent of noise power such that (\ref{t4}) does
not have noise uncertainty. Thus, they are different. Note that the
new detectors do use the extra knowledge of the interval and the
distribution of the noise power. This can be considered as a
stochastic maximum likelihood method when the unknown parameter is
eliminated by averaging over its distribution. The values of $\Delta
_{max}$ and $\Delta _{min}$ are easier to obtain than $\sigma^2$, as they are
only determined by the upper and lower limits of the possible range
of $\sigma ^2$. In practice, they can be calculated from the
operating range of the receiver temperature when the bandwidth is
fixed. Note also that all the detectors are compared based on the
assumption of independent samples. This assumption has been widely
used in the literature \cite{Stuber}. For the noise samples, this
can be achieved by Nyquist sampling. For the signal samples, this
can be achieved when the Doppler shift is large or the sampling
interval is large.

%%%%%%%%%%%%%%%%%%%%%%%%%%%%%%%%%%%%%%%%%%%%%%%%%%%%%%%%%%%%
\section{Numerical results and discussion}
In this section, the performances of the conventional detector with the
maximum likelihood estimate given in Section
2 and the three new detectors
derived in Section 3 are
evaluated via computer simulation. Define the received signal power
as $P=\beta ^2$ and assume that the noise power $\sigma ^2$ is
uniformly distributed over the interval $\left (\Delta_{min},
\Delta_{max} \right )$. Define $SNR=\frac{2
\;P}{\Delta_{max}+\Delta_{min}}$. In all the figures, ``NP-LRT''
refers to the conventional detector in (\ref{eqe2}), ``NP-AVE''
refers to the new detector in (\ref{eq13}), ``NP-AVN'' corresponds
to the new detector in (\ref{eq19}) and ``NP-LLR'' refers to the new
detector in (\ref{t4}). The thresholds $\gamma_2$ and $\gamma_3$ are
calculated from $10^6$ Monte Carlo trials using the NP rule while
the threshold $\gamma_1$ and the threshold $\gamma_4$ are calculated
by (\ref{eqq7}) and (\ref{eqe9}), respectively. Also, assume that
the bandwidth is $B=6 MHz$ and $K=4*10^{13}$, together with the
Boltzmann constant $k=1.38*10^{-23}$. Then, the noise power of
$0.5$, $0.7$, $1.3$, $1.5$ in the simulation below correspond to the
temperature of 150K, 210K, 391K, 451K, respectively.

\begin{figure}[htbp]
\centering \subfigure[]{\includegraphics [width=4.5in]{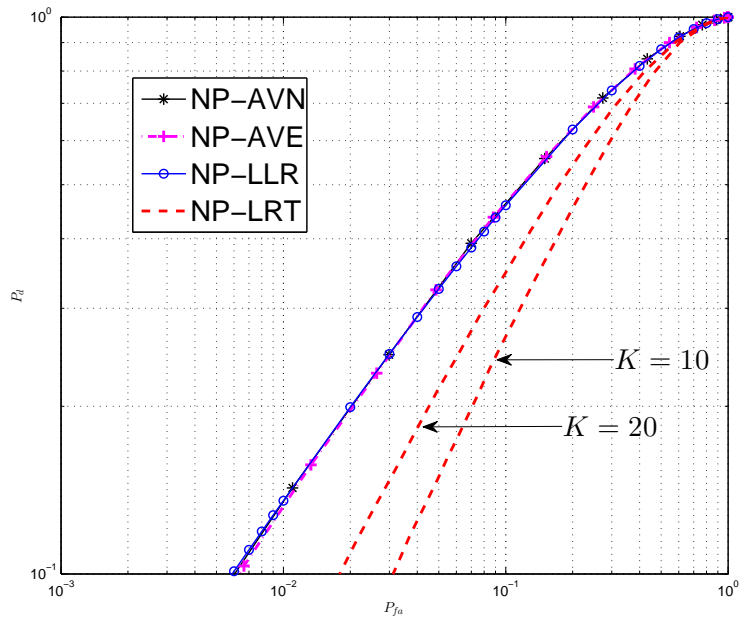}}
\subfigure[]{\includegraphics [width=4.5in]{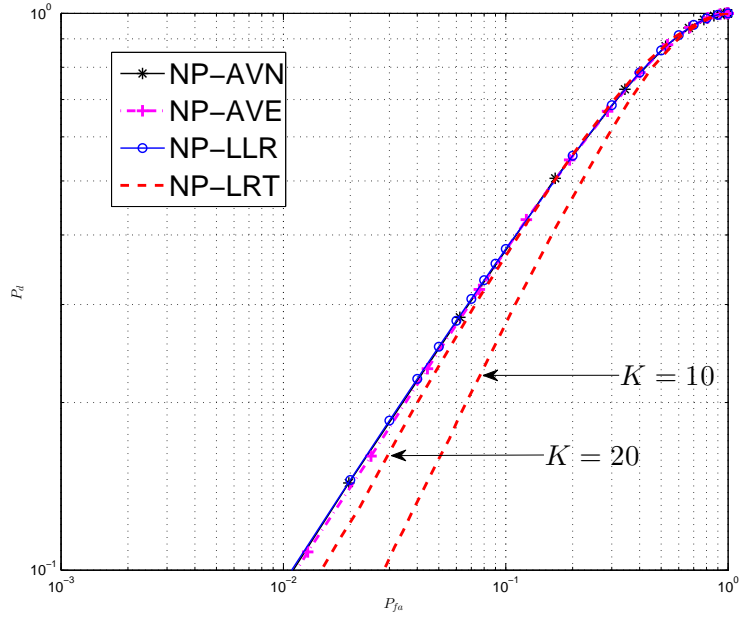}} \caption{ROC
curves for $P = 0.5$ and $N = 20$. (a) $\Delta _{min}$ = 0.7 and
$\Delta _{max}$ = 1.3. (b) $\Delta _{min}$ = 0.5 and $\Delta
_{max}$=1.5.} \label{fig1}
\end{figure}

Fig. \ref{fig1} compares the ROC curves for the NP-AVN, NP-AVE,
NP-LLR and NP-LRT using maximum likelihood estimate with $K=10$ or
$K=20$ at $P = 0.5$ and $N = 20$. Fig. \ref{fig1}(a) considers the
noise interval as $\Delta _{min}$ = 0.7 and $\Delta _{max}$ = 1.3
while Fig. \ref{fig1}(b) enlarges the noise interval to $\Delta
_{min}$ = 0.5 and $\Delta _{max}$ = 1.5. One can see that the
performances of these four detectors degrade when the noise interval
increases, as expected. On the other hand, the performance gains of
the new detectors over the conventional detector are large but
decrease when the noise interval increases.

\begin{figure}[htbp]
\centering \subfigure[]{\includegraphics [width=4.5in]{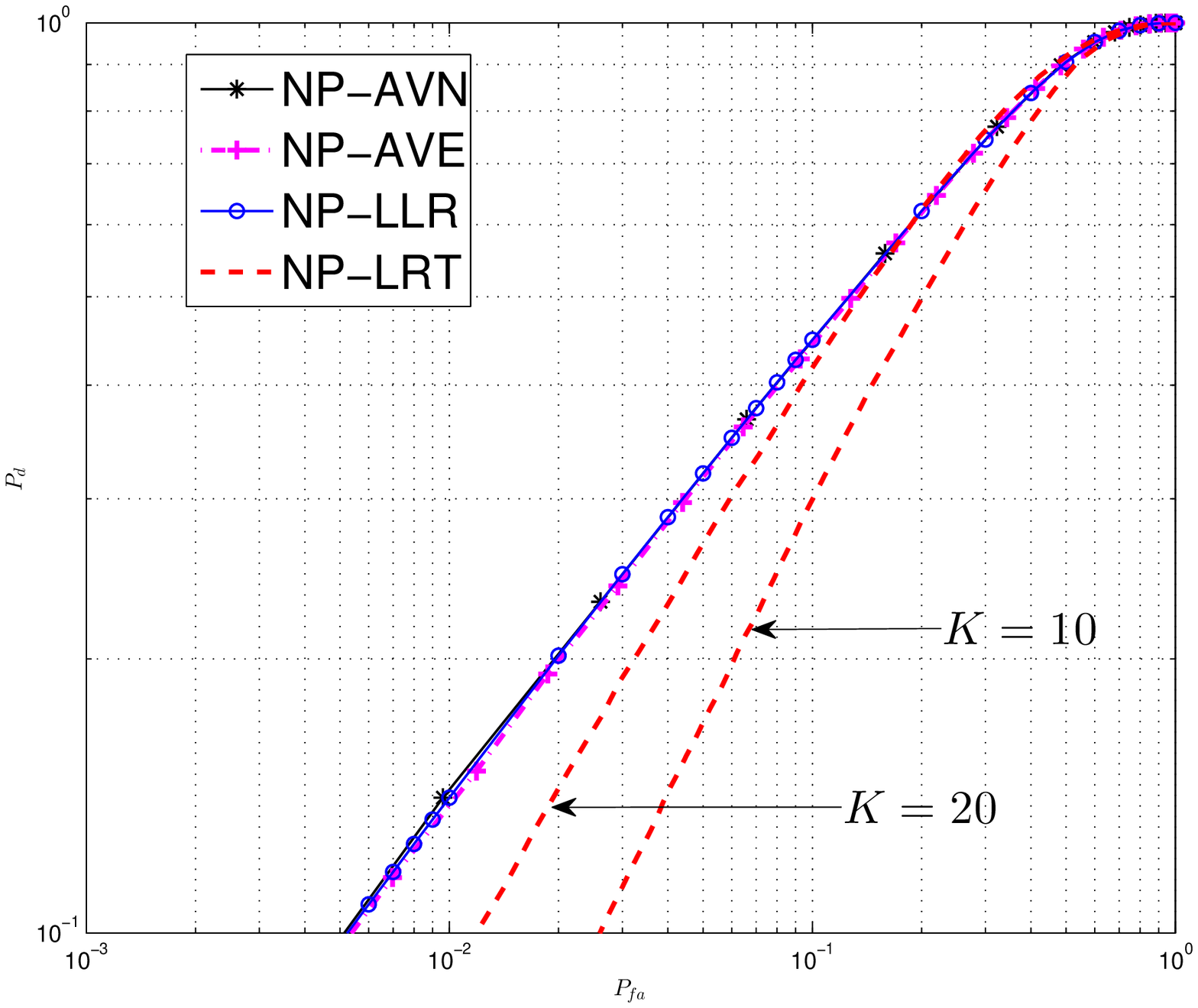}}
\subfigure[]{\includegraphics [width=4.5in]{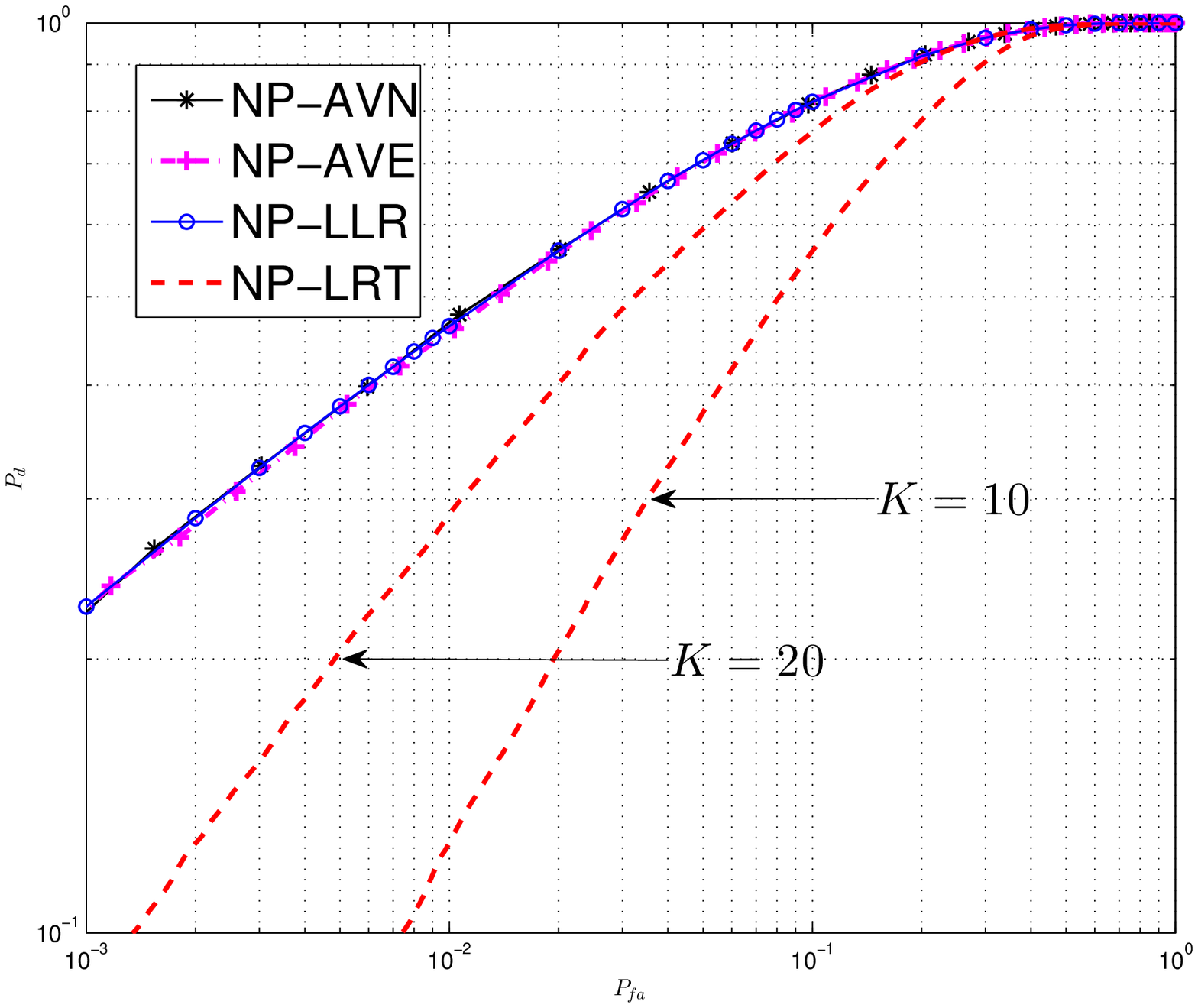}} \caption{ROC
curves for $N = 40$, $\Delta _{min}$ = 0.5 and $\Delta _{max}$ =
1.5. (a) $P = 0.5$. (b) $P = 1$.} \label{fig2}
\end{figure}

Fig. \ref{fig2} compares the ROC curves for the NP-AVN, NP-AVE,
NP-LLR and NP-LRT with $K=10$ or $K=20$ at $N = 40$, $\Delta _{min}$
= 0.5 and $\Delta _{max}$ = 1.5 for different $P$. One can see that
the performance gains of these new detectors increase when the
received signal power increases. Also, one can see from Fig.
\ref{fig1}(b) and Fig. \ref{fig2}(a) that the performance gains of
the new detectors over the conventional detector increase when $N$
increases. The performance gains of the new detectors over the
conventional detector are still substantial even at low SNRs.

\begin{figure}[htbp]
\centering \subfigure[]{\includegraphics [width=4.5in]{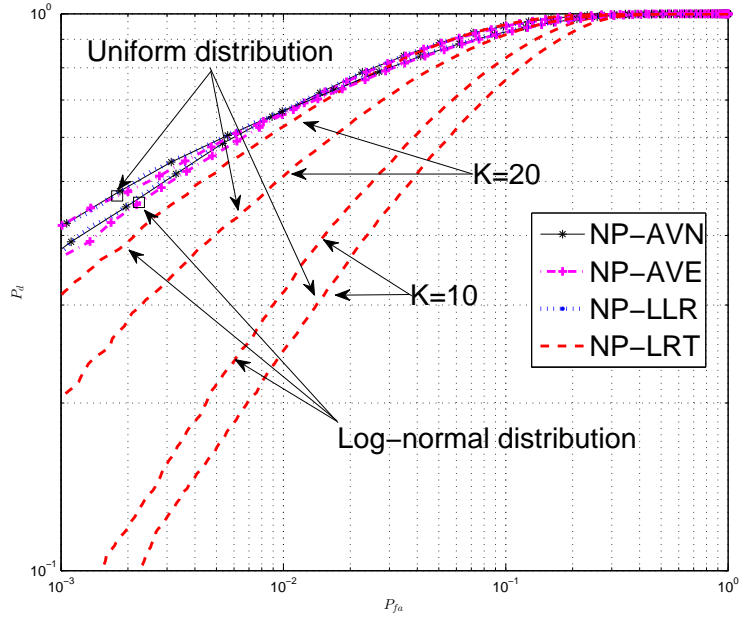}}
\subfigure[]{\includegraphics [width=4.5in]{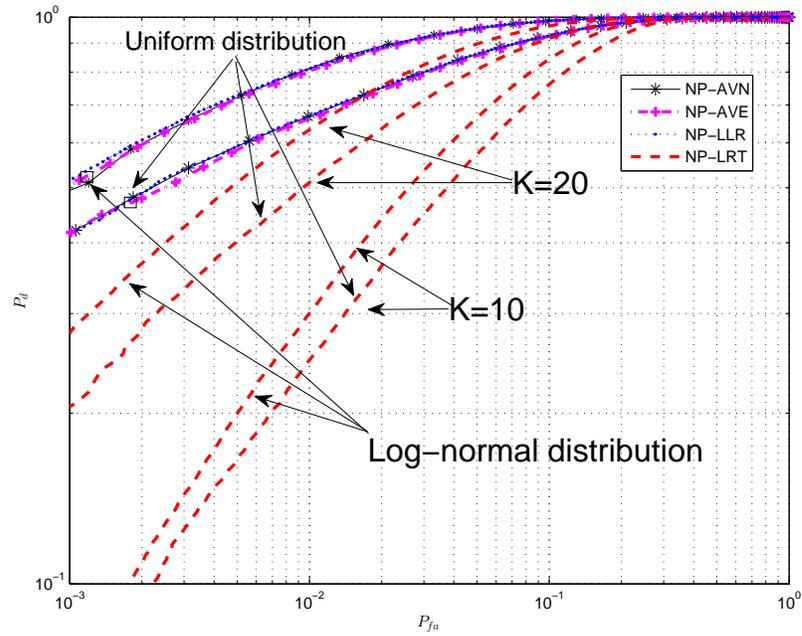}} \caption{ROC curves for $P = 1.8$, $N = 40$, $\Delta _{min} =
\frac{1}{2}$ and $\Delta _{max} = 2$ when noise power follows a
log-normal distribution with variance $\sigma^2_n$ while
assuming a uniform distribution. (a) $\sigma^2_n = 1$ (b)$\sigma^2_n = 0.1$. } \label{fig3}
\end{figure}

\begin{comment}
Fig. \ref{ff} compares $P_d$ vs $SNR$ for NP-AVN, NP-AVE, NP-LLR and
NP-LRT when $N = 20$, $\Delta _{min}$ = 0.7, $\Delta
_{max}$ = 1.3 and $P_{fa}=0.1$. One can see the improvement of the
three new detectors in $P_d$ is most significant at low SNRs, as is
the case in reality.
\begin{figure}[htbp]
\centering
\includegraphics[width=3in]{5.eps}
\caption{$P_d$ vs $SNR$ at $N = 20$, $\Delta _{min}$ =
0.7 and $\Delta _{max}$ = 1.3 when $P_{fa}=0.1$.} \label{ff}
\end{figure}
\end{comment}
Fig. \ref{fig3} compares the ROC curves for the NP-AVN, NP-AVE,
NP-LLR and NP-LRT at $P = 1.8$ and $N = 40$ when the noise power
follows a log-normal distribution with variance $\sigma^2_n$ and the uniform distribution in the interval between $\Delta _{min} =
\frac{1}{2}$ and $\Delta _{max} = 2$. Fig. \ref{fig3} (a) considers $\sigma^2_n=1$ while Fig. \ref{fig3} (b) reduce the variance to $\sigma^2_n=0.1$.
These figures are used to examine
the effect of mismatch between assumed and actual noise power
distributions on the performances of the new detectors, as the
simulated samples are generated using log-normal distribution while
the derivation in Section 3
assumes a uniform distribution. One can see that the three new
detectors based on the uniform distribution still have considerable
gains over the conventional detector even when the actual noise
power follows a log-normal distribution.
Moreover in Fig. \ref{fig3} (a), one can see that
the
performances of the new detectors do degrade for small values of
$P_{fa}$ when there is a mismatch. However the performance
degradation is quite small compared to their performance gains over
the conventional detector. On the other hand, when the variance of log-normal distribution decreases, the gain of our new detectors over the conventional one increases.

\begin{figure}[htbp]
\centering
\includegraphics[width=4.5in]{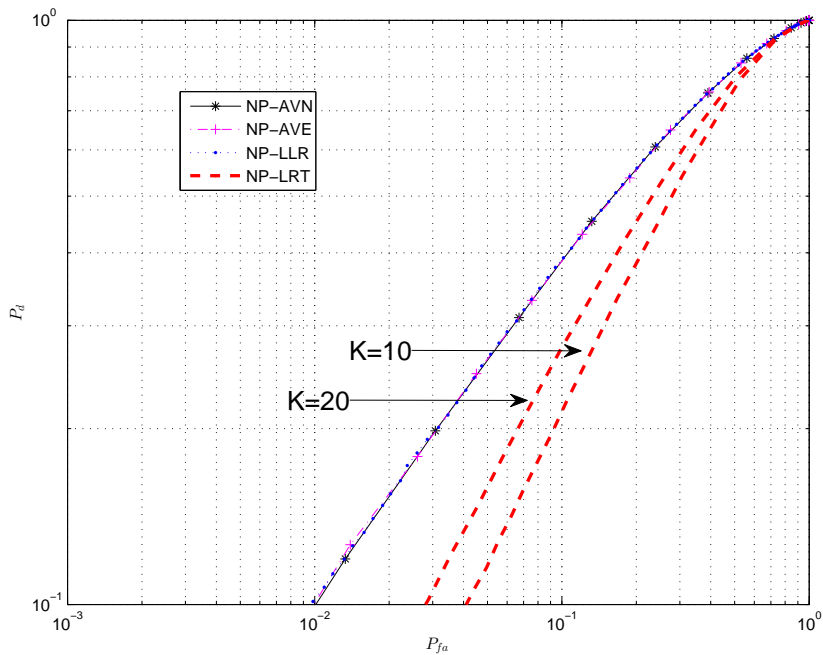}
\caption{ROC
curves for $P = 0.5$, $N = 20$, $\Delta _{min}$ = 0.7 and
$\Delta _{max}$ = 1.3 when the power of interference is assumed $\eta=0.3$.} \label{fig4}
\end{figure}

Fig. \ref{fig4} compares the ROC curves for the NP-AVN, NP-AVE,
NP-LLR and NP-LRT at $P = 0.5$, $N = 20$, $\Delta _{min} =
0.7$ and $\Delta _{max} = 1.3$ when the interference is assumed to
follow a normal distribution with variance $\eta=0.3$.
This figure is used to examine
the performances of new detectors over the conventional detector when the interference exists.
Comparing Fig. \ref{fig4} with Fig. \ref{fig1} (a), one can see that, although the performances of new detectors degrade, there is still considerable gain of the new detectors over the conventional one in Fig. \ref{fig4}. Therefore, our new detectors are still useful even when interference exists.

As expected, one can see from these figures that the performance of
NP-LRT detector improves when the number of samples increases. Also,
one can find the NP-LLR detector gives the best detection
performance while the conventional NP-LRT detector has the worst
performance. Although the NP-AVN, NP-AVE and NP-LLR detectors have
very close performances, they have different structures and
complexities. NP-AVN detector has a slightly better performance but
a more complicated structure than the NP-AVE detector because it
includes exponential integral function. Specifically, NP-AVN
detector takes 0.54 seconds while NP-AVE detector only takes 0.001
seconds in Matlab R2013a simulation using a computer with 64-bit
operation system, CPU i7-2600 and 4G memory. Moreover, the NP-LLR
detector has closed-form expressions for the threshold, the
detection probability and the false alarm probability. Thus, one can
choose the suitable detector according to their application for
different performances or complexities. We have also found that the
NP-AVN detector is more robust than NP-AVE detector to the mismatch between the assumed noise power interval and
the true noise power interval.

%Due to the length
%restriction, these simulation results are not shown here.
\section{Conclusion}
New energy detectors based on the uniform distribution of the noise
power have been proposed. Numerical results have shown that these
new detectors outperform the conventional detector in the presence
of noise uncertainty. The performance gain depends on the received
signal power, the noise power interval and the number of samples.
This gain is achieved by using the extra knowledge of the uniform
distribution and comparison of these detectors reveals the effect of
this extra knowledge and therefore is useful. The new detectors however have similar performances, as can be seen from the figures.

{\section*{APPENDIX A: DERIVATION OF THE NP-AVE DETECTOR}} In this
case, the averaged likelihood function under hypothesis $H_1$
becomes
\begin{equation}\label{eq9}
f(x[i]|H_1)=\int _{\Delta_{min}}^{\Delta_{max}}f(x[i]| H_1,\sigma
^2)f\left(\sigma ^2\right) d\sigma ^2,
\end{equation}
and the averaged likelihood function under $H_0$ becomes
\begin{equation}\label{eq11}
f(x[i]| H_0)=\int _{\Delta _{min}}^{\Delta _{max}}f(x[i]| H_0,\sigma
^2)f\left(\sigma ^2\right) d\sigma ^2.
\end{equation}
Using the independence of samples, $\text{Erf}(z)=\frac{2}{\sqrt{\pi
}}\int _0^ze^{-t^2}dt$ \cite{Gradshteyn} and equation (\ref{pdf}),
one can get (\ref{eq10}) and (\ref{eq12}).

%\subsection{Derivation of the  NP-AVN detector}
{\section*{APPENDIX B: DERIVATION OF THE NP-AVN DETECTOR}} In this
case, by averaging the joint PDF of all samples over $\sigma ^2$
under hypothesis $H_1$, (\ref{eq6}) becomes
\begin{equation}\label{eq15}
f(X|H_1)=\int _{\Delta _{min}}^{\Delta _{max}}f\left(X|H_1,\sigma
^2\right)f\left(\sigma ^2\right) d\sigma ^2
\end{equation}
which gives
\begin{equation}\label{eq16}
\begin{aligned}
&f(X|H_1)=\frac{(2 \pi )^{-N/2} }{\left(\Delta _{max}-\Delta
_{min}\right)} \left(\left(\frac{\sum _{i=1}^N
x[i]^2}{2}\right)^{1-N/2}\text{
}\Gamma\left(\frac{N}{2}-1,\frac{\sum _{i=1}^N x[i]^2}{2
\left(\beta ^2+\Delta _{max}\right)}\right)-\right. \\
&\left.\left(\frac{\sum _{i=1}^N x[i]^2}{2} \right)^{1-N/2}\text{
}\Gamma\left(\frac{N}{2}-1,\frac{\sum _{i=1}^N x[i]^2 }{2
\left(\beta ^2 +\Delta _{min}\right)}\right)\right)
\end{aligned}
\end{equation}
where the incomplete gamma function is given by $\Gamma (a,z)=\int
_z^{\infty }t^{a-1}e^{-t}dt$ \cite{Gradshteyn}.

Similarly, under hypothesis $H_0$, the averaged likelihood function
is
\begin{equation}\label{eq17}
f(X|H_0)=\int _{\Delta _{min}}^{\Delta _{max}}f\left(X|H_0,\sigma
^2\right)f\left(\sigma ^2\right) d\sigma ^2
\end{equation}
which gives
\begin{equation}\label{eq18}
\begin{aligned}
&f(X|H_0)=\frac{(2 \pi )^{-N/2}}{\left(\Delta _{max}-\Delta
_{max}\right)} \left(\left(\frac{\sum _{i=1}^N
x[i]^2}{2}\right)^{1-\frac{N}{2}}
\times\Gamma\left(\frac{N}{2}-1,\frac{\sum _{i=1}^N x[i]^2}{2
\Delta _{max}}\right)\right. \\
&\left.-\left(\frac{\sum _{i=1}^N
x[i]^2}{2}\right)^{1-\frac{N}{2}}\text{
}\Gamma\left(\frac{N}{2}-1,\frac{\sum _{i=1}^N x[i]^2 }{2 \Delta
_{min}}\right)\right).
\end{aligned}
\end{equation}
Then, the likelihood ratio test is given as
\begin{equation}\label{eqq19}
\begin{aligned}
&L_3(X)= \frac{f(X|H_1)}{ f(X|H_0)}
\end{aligned}
\end{equation}
By using the independence of different samples,
$\text{\textit{$EI$}}(\text{\textit{$n$}},\text{\textit{$z$}})=z^{n-1}
\Gamma (1-n,z)$ \cite{Gradshteyn} together with (\ref{eq16}) and
(\ref{eq18}), one can get (\ref{eq19}).

{\section*{APPENDIX C: DERIVATION OF THE NP-LLR DETECTOR}}
 The PDF
of $L_4(X)$ under $H_0$ can be shown to follow a chi-square
distribution as \cite{Kay}
\begin{equation}\label{eqe3}
\begin{aligned}
&f_{L_4|H_0}(X | \sigma ^2)=\\&\frac{2^{-\frac{N}{2}} \sigma^{-N}
(\sum _{i=1}^N x[i]^2)^{\frac{N}{2}-1} e^{-\frac{\sum _{i=1}^N
x[i]^2}{2 \sigma ^2}}}{\Gamma \left(\frac{N}{2}\right)}.
\end{aligned}
\end{equation}
Using (\ref{pdf}) and (\ref{eqe3}), one can get
\begin{equation}\label{eqe4}
\begin{aligned}
&f_{L_4|H_0}(X)=\int _{\Delta _{min}}^{\Delta _{max}}f_{L_4|H_0}(X
|\sigma ^2)f\left(\sigma ^2\right)d\sigma
^2\\&=\left(2^{-\frac{N}{2}} (\sum _{i=1}^N x[i]^2)^{\frac{N}{2}-1}
\left(2^{\frac{N}{2}-1} \sum _{i=1}^N x[i]^2 \Delta
_{max}^{-\frac{N}{2}} \left(\frac{\sum _{i=1}^N x[i]^2}{\Delta
_{max}}\right)^{-\frac{N}{2}} \Gamma \left(\frac{N-2}{2},\frac{\sum
_{i=1}^N x[i]^2}{2 \Delta _{max}}\right)\right.\right. \\
&\left.\left.-2^{\frac{N}{2}-1} \sum _{i=1}^N x[i]^2 \Delta
_{min}^{-\frac{N}{2}} \left(\frac{\sum _{i=1}^N x[i]^2}{\Delta
_{min}}\right)^{-\frac{N}{2}}\times
\Gamma \left(\frac{N-2}{2},\frac{\sum _{i=1}^N x[i]^2}{2 \Delta
_{min}}\right)\right)\right)\\&/\left((\Delta _{max}-\Delta _{min})
\Gamma \left(\frac{N}{2}\right)\right).
\end{aligned}
\end{equation}
Following the same definition as (\ref{eqe5}) and (\ref{eqe6}), one
can get $P_{fa}$ as
\begin{equation}\label{e7}
\begin{aligned}
&P_{fa}=Pr\{L_4(X)>\gamma_4|H_0\}= \int
_{\gamma_4}^{\infty}f_{L_4|H_0}(X)dX,
\end{aligned}
\end{equation}
and $P_d$ as
\begin{equation}\label{e8}
\begin{aligned}
&P_d=Pr\{L_4(X)>\gamma_4|H_1\}=\int
_{\gamma_4}^{\infty}f_{L_4|H_1}(X)dX.
\end{aligned}
\end{equation}
After simplifications, one can get (\ref{eqe7}) and (\ref{eqe8}).

\bibliographystyle{wileyj}
\bibliography{wcmdoc}

\end{document}